\documentclass{emulateapj}

\received{February 6, 2004}
\accepted{March 19, 2004}

\newcommand{\pcmsq}{\mbox{cm$^{-2}$}}
\newcommand{\cmsq}{\mbox{cm$^{2}$}}
\newcommand{\ergsec}{\mbox{erg s$^{-1}$}}

\newcommand{\Lx}{\mbox{$L_{\rm x}$}}

\newcommand{\Msun}{\mbox{M$_\odot$}}

\newcommand{\nh}{\mbox{N$_H$}}
\newcommand{\chandra}{\textit{Chandra}}
\newcommand{\hst}{\textit{Hubble Space Telescope}}
\newcommand{\cxo}{\textit{Chandra X-ray Observatory}}
\newcommand{\rosat}{\textit{ROSAT}}
\newcommand{\xspec}{\textsc{XSPEC}}
\newcommand{\tool}{\it}
\newcommand{\xsoft}{\mbox{X$_{\rm soft}$}}
\newcommand{\xmed}{\mbox{X$_{\rm med}$}}
\newcommand{\xhard}{\mbox{X$_{\rm hard}$}}

\newcommand{\err}[2]{\small \ensuremath{^{+#1}_{-#2}}}
\newcommand{\ee}[2]{\ensuremath{#1\times 10^{#2}}}
\newcommand{\filt}[2]{\ensuremath{{#1}_{#2}}}
\def\lapp{\ifmmode\stackrel{<}{_{\sim}}\else$\stackrel{<}{_{\sim}}$\fi}

\shorttitle{X-ray Sources in M4}
\shortauthors{Bassa et al.}
\slugcomment{}

\begin{document}
  \title{X-ray Sources and their Optical Counterparts in the Globular
    Cluster M4}
  \author{Cees~Bassa\altaffilmark{1}, 
    David~Pooley\altaffilmark{2},
    Lee~Homer\altaffilmark{3},
    Frank~Verbunt\altaffilmark{1},
    Bryan~M.~Gaensler\altaffilmark{4},
    Walter~H.~G.~Lewin\altaffilmark{2},
    Scott~F.~Anderson\altaffilmark{3},
    Bruce~Margon\altaffilmark{5},
    Victoria~M.~Kaspi\altaffilmark{6,2} and
    Michiel~van~der~Klis\altaffilmark{7}
  }

  \altaffiltext{1}{Astronomical Institute, Utrecht University, PO Box
    80\,000, 3508~TA, Utrecht, The Netherlands; c.g.bassa@astro.uu.nl,
    f.w.m.verbunt@astro.uu.nl.}
  \altaffiltext{2}{Center for Space Research and Department of Physics,
    Massachusetts Institute of Technology, 70 Vassar Street, Building
    37, Cambridge, MA 02139-4307; davep@space.mit.edu, lewin@space.mit.edu.}
  \altaffiltext{3}{Department of Astronomy, University of
    Washington, Box 351580, Seattle, WA 98195-1580;
    homer@astro.washington.edu, anderson@washington.edu.}
  \altaffiltext{4}{Harvard-Smithsonian Center for Astrophysics, 60
    Garden Street, Cambridge, MA 02138; bgaensler@cfa.harvard.edu.}
  \altaffiltext{5}{Space Telescope Science Institute, 3700 San Martin
    Drive, Baltimore, MD 21218; margon@stsci.edu}
  \altaffiltext{6}{Department of Physics, McGill University, Ernest
    Rutherford Physics Building, 3600 University Street, Montreal, QC
    H3A 2T8, Canada; vkaspi@hep.physics.mcgill.ca}
  \altaffiltext{7}{Astronomical Institute ``Anton Pannekoek'',
    University of Amsterdam, Kruislaan 403, 1098 SJ, The Netherlands;
    michiel@science.uva.nl}

  \begin{abstract}
    We report on the \cxo\ ACIS-S3 imaging observation of the Galactic
    globular cluster M4 (NGC\,6121). We detect 12 X-ray sources inside
    the core and 19 more within the cluster half-mass radius. The
    limiting luminosity of this observation is
    $\Lx\approx10^{29}$~\ergsec\ for sources associated with the
    cluster, the deepest X-ray observation of a globular cluster to
    date. We identify 6 X-ray sources with known objects and use
    \rosat\ observations to show that the brightest X-ray source is
    variable. Archival data from the \emph{Hubble Space Telescope}
    allow us to identify optical counterparts to 16 X-ray
    sources. Based on the X-ray and optical properties of the
    identifications and the information from the literature, we
    classify two (possibly three) sources as cataclysmic variables,
    one X-ray source as a millisecond pulsar and 12 sources as
    chromospherically active binaries. Comparison of M4 with 47\,Tuc
    and NGC\,6397 suggests a scaling of the number of active binaries in
    these clusters with the cluster (core) mass.
  \end{abstract}


  \section{Introduction}
  Faint X-ray sources ($\Lx \lapp 10^{35}$~\ergsec) were first
  discovered in observations made with the \textit{Einstein} and
  \rosat\ observatories (Hertz \& Grindlay\,1983\nocite{hg83};
  Verbunt\,2001\nocite{ver01}). It was only with \chandra, however,
  that a large number of faint sources was identified: more than one
  hundred in 47\,Tuc, and up to a few dozen each in e.g.\ NGC\,6397,
  NGC\,6752, NGC\,6440, and $\omega$\,Cen (Grindlay et
  al.\,2001a,b\nocite{ghem01,ghe+01}; Pooley et
  al.\,2002a,b\nocite{plh+02,plv+02}; Rutledge et
  al.\,2002\nocite{rbb+02}). These faint sources represent a mix of
  objects with different X-ray luminosities. Brightest are the neutron
  stars accreting at a low rate from a companion (quiescent low mass
  X-ray binaries or qLMXBs), followed by white dwarfs accreting from
  low-mass companions (cataclysmic variables or CVs). The active
  binaries (ABs) tend to be the faintest, while radio pulsars with
  short periods (millisecond or recycled pulsars, MSPs) have X-ray
  luminosities in a similar range as CVs. The three types of
  chromospherically or magnetically active binaries are detached
  binaries of two main sequence stars (BY~Dra systems), detached
  binaries of a main sequence star and a giant or a sub-giant (RS~CVn
  systems) and contact binaries (W~UMa systems).

  Globular clusters contain many more neutron star binaries per unit
  mass than the galactic disk. Hence, if in a globular cluster one
  finds a binary with a neutron star (which is rather difficult to
  make via ordinary evolution of an initial binary), it is highly
  probable that this binary was formed via a close encounter between
  stars (Fabian et al.\,1975\nocite{fpr75};
  Hills\,1976\nocite{hil76}). Isolated millisecond pulsars in globular
  clusters are thought to have been formed in such binaries, and thus
  are probably also a result of close stellar encounters.

  On the other hand, binaries that are common in the field, are more
  likely to be of primordial origin in globular cluster, including the
  chromospherically active binaries. Cataclysmic variables are
  relatively common in the Galactic disk, and those in globular
  clusters could in principle originate from primordial binaries or
  stellar encounters. Pooley et al.\,(2003\nocite{pla+03}) showed that
  the number of faint sources above the threshold $\Lx>
  4\times10^{30}$ \ergsec\ in a cluster scales with its collision
  number $\Gamma$, which is a theoretical estimate for the number of
  close encounters. Since the majority of such sources are cataclysmic
  variables, this suggests that most cataclysmic variables in globular
  clusters are in fact formed via close stellar encounters.

  Most clusters investigated with \chandra\ so far have relatively
  high collision numbers (see e.g.\ Table~1 in Pooley et
  al.\,2003\nocite{pla+03}); because the limit to which X-ray sources
  in these clusters could be detected is relatively high, most X-ray
  sources known in these clusters are either neutron stars (accreting
  or radio pulsars), or cataclysmic variables. So far, a sizable
  number of X-ray sources has been identified as chromospherically
  active binaries only in 47\,Tuc (Edmonds et
  al.\,2003a,b\nocite{eghg03a,eghg03b}). In this paper, we discuss a
  globular cluster with a relatively low collision number, M4
  (NGC\,6121). This cluster is a relatively nearby cluster with a
  moderate absorption ($d=1.73$~kpc, $A_V=1.32$, Richer et
  al.\,1997\nocite{rfi+97}). The core and half-mass radii of M4 are
  $49\farcs8$ and $3\farcm65$, respectively
  (Harris\,1996\nocite{har96}). We use these values throughout the
  paper.

  A priori, we would thus expect that the majority of X-ray sources in
  this cluster to be chromospherically active binaries. That
  chromospherically active binaries exist in this cluster is evident
  from optical studies.  For example, Kaluzny et
  al.\,(1997\nocite{ktk97}) discovered a number of optical variables,
  including several contact binaries.

  The presence of a recycled radio pulsar PSR~B1620$-$26 (Lyne et
  al.\,1988\nocite{lbb+88}) in M4 is remarkable, but can be explained
  by noting that the small collision numbers of many globular clusters
  still add up, so that at least some of the clusters with small
  $\Gamma$ should contain a binary with a neutron star (Verbunt \&
  Hut\,1987\nocite{vh87}). PSR~B1620$-$26 is in a 191 day orbit around
  a white dwarf of $\sim\!0.3$~\Msun; the binary is accompanied by a
  third object of planetary mass in an orbit of $\sim100$~years
  (Thorsett et al.\,1999\nocite{tacl99}; Sigurdsson et
  al.\,2003\nocite{srh+03}).

  \section{X-ray Observations and Analysis}
  M4 was observed for 25.8~ks on 2000~June~30 with the Advanced CCD
  Imaging Spectrometer (ACIS) on the \cxo\ with the telescope aimpoint
  on the back-side illuminated S3 chip. The data were taken in
  timed-exposure mode with the standard integration time of 3.24~s per
  frame and telemetered to the ground in faint mode.

  Data reduction was performed using the CIAO~2.3 software provided by
  the \chandra\ X-ray Center
  (CXC)\footnote{\url{http://asc.harvard.edu}}.  We reprocessed the data
  using the CALDB\,2.21 set of calibration files (gain maps, quantum
  efficiency, quantum efficiency uniformity, effective area) without
  including the pixel randomization that is added during standard
  processing.  This method slightly improves the point spread function.
  We filtered the data using the standard {\it ASCA} grades, and we
  excluded both bad pixels and software-flagged cosmic ray
  events. Intervals of background flaring were searched for, but none
  were found, hence we simply applied the good-time intervals supplied
  by the CXC.

  \begin{figure}
    \resizebox{85mm}{!}{\plotone{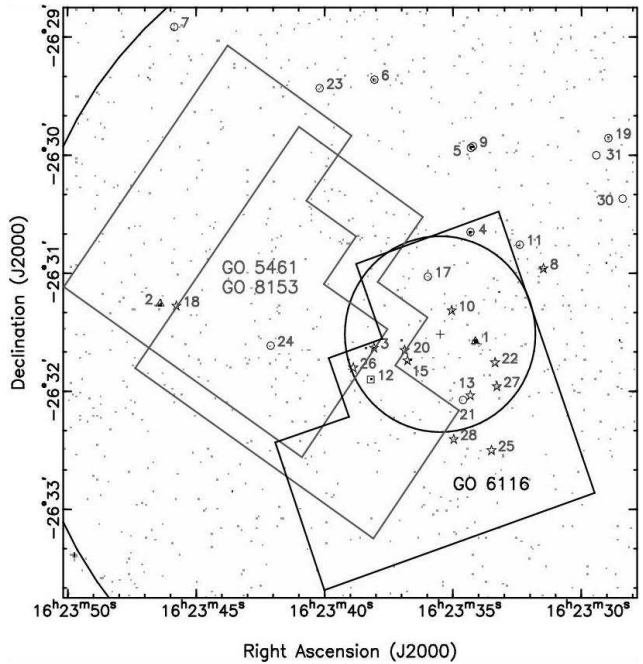}}
    \caption{X-ray image of a $5\arcmin\times5\arcmin$ region of
      M4. Shown are all counts in the 0.5--6.0~keV range. The cluster
      center, core and half-mass radii are indicated by a cross and
      two circles, respectively (Harris\,1996). The detected X-ray
      sources are marked and numbered, omitting the ``CX''
      prefix. X-ray sources classified as cataclysmic variables are
      indicated with triangles, chromospherically active binaries with
      stars, millisecond pulsars with squares and unclassified or
      ambiguous sources are indicated with circles. See
      Section~\ref{sec:classify} for the source
      classification. Also shown are three archival pointings of the
      \emph{HST}/WFPC2 used to locate optical counterparts to the
      \chandra\ X-ray sources.\label{fig:xrayimage}}
  \end{figure}

  \subsection{Source Detection} \label{sect:detection}
  The CIAO wavelet-based {\tool wavdetect} tool was employed for
  source detection in both the 0.5--6.0~keV band and 0.3-10.0~keV band.
  We detected 41 sources on the entire S3 chip in the 0.5--6.0~keV band;
  for the {\tool wavdetect} settings we used, approximately one of
  those detections may be spurious.  Of these sources, 30 lie within
  the half-mass radius of the cluster.

  We also searched part of the adjacent S4 CCD since part of the
  half-mass region fell on this chip, but no sources were detected in
  this area.  One additional source was detected on S3 in the broader
  0.3--10.0~keV band with three counts, two of which were between 8
  and 10~keV.  The significance of this source, as well as its
  possible membership of the cluster, is questionable.  For
  completeness, however, we leave it in our source list
  (Table~\ref{tab:srcs}) as the last source.  The sources are numbered
  according to detected counts in the 0.5--6.0~keV band.

  From the $\log{N}-\log{S}$ relationships of Giacconi et
  al.\,(2001\nocite{grt+01}), we expect between 5 and 6 background
  sources within the cluster half-mass radius. However, as M4 is
  located towards the bulge of the galaxy ($l=351.0\degr$,
  $b=16.0\degr$, Harris\,1996\nocite{har96}) this relation would
  underestimate the galactic contribution of background sources and
  provide a lower limit only. We have analyzed ACIS-S3 data from the
  October 15, 2002 \chandra\ observation of the low-mass X-ray binary
  (LMXB) MXB~1659$-$298, which has $l=353.8\degr$ and $b=7.3\degr$. In
  the 27~ks of this observation the 1/4 sub-array of the S3 CCD shows 5
  sources, other than the LMXB itself, in the 0.5--6.0~keV band. Scaling
  this up to the whole chip and taking into account Poissonian errors
  would give rise to some 20 sources within an area of the size of the
  M4 half-mass radius. Given that MXB~1659$-$298 is located about
  $10\degr$ closer to the galactic center this would overestimate the
  number of background sources towards M4. Though both limit
  determinations are very different in nature, we expect 6 to 20
  background sources within the half-mass radius of M4, and zero or
  one within the core.
 
  All sources are consistent with being point sources.  However, CX5 and
  CX9 overlap somewhat.  They were identified by {\tool wavdetect} as
  two separate sources.  It is possible that they are actually a single
  extended object, but the double-peaked nature of the image lends
  support to the interpretation as two point sources.

  There is some evidence for excess emission on the approximately
  32\,200 pixels inside the core radius. The residual number of counts
  (i.e., the total counts minus those due to the 12 point sources
  inside the core) in the 0.5--6.0~keV band is $438\pm20$~counts,
  while we would expect $287\pm15$ background counts on the basis of
  the number of counts that we measure in a source-free region outside
  the core. This excess emission could be due to either unresolved
  point sources or diffuse emission.

  A $5\arcmin\times5\arcmin$ portion of the \chandra\ exposure
  of M4 is shown in Figure~\ref{fig:xrayimage}. The detected
  sources are encircled and numbered. The cross indicates the cluster
  center, while the circles denote the core and half-mass radius. 

  \begin{deluxetable*}{lllcccccc}
    \tabletypesize{\footnotesize}
    \tablecolumns{9}
    \tablewidth{0pc}
    \tablecaption{M4 X-Ray Sources\label{tab:srcs}}
    \tablehead{
      \colhead{} & \colhead{} & \colhead{} &
      \multicolumn{3}{c}{Detected Counts/Corrected Counts\tablenotemark{c}} & 
      \colhead{} & \colhead{} & \colhead{} \\
      \colhead{} & \colhead{R.A.} & \colhead{Decl.} &
      \colhead{} & \colhead{} & \colhead{} & 
      \colhead{\Lx (0.5--2.5~keV)} & \colhead{} & \colhead{} \\
      \colhead{Source\tablenotemark{a}} &
      \colhead{(J2000.0)\tablenotemark{b}} &
      \colhead{(J2000.0)\tablenotemark{b}} & 
      \colhead{$\mathrm{X}_\mathrm{soft}$} &
      \colhead{$\mathrm{X}_\mathrm{med}$} &
      \colhead{$\mathrm{X}_\mathrm{hard}$} &
      \colhead{(ergs\ s$^{-1})$\tablenotemark{d} } &
      \colhead{Counterpart\tablenotemark{e}} &
      \colhead{ID\tablenotemark{f}}
    }
    \startdata
    CX 1\ldots 	&  16 23 34.128 (1) 	& -26 31 34.85 (2) 	& 154/381.2 	& 379/736.4 	& 275/313.0 	& $8.3\times10^{31}$ 	& Opt.?/R9?	& CV \\
    CX 2\ldots 	&  16 23 46.399 (3) 	& -26 31 15.67 (8) 	&  48/117.5 	&  72/138.3 	&  30/33.8  	& $6.0\times10^{30}$ 	& Opt.	        & CV \\
    CX 3\ldots 	&  16 23 38.073 (3) 	& -26 31 38.18 (4) 	&  46/113.6 	&  75/145.2 	&  29/32.8  	& $5.4\times10^{30}$ 	& Opt.	        & AB \\
    CX 4\ldots 	&  16 23 34.321 (4) 	& -26 30 39.27 (5) 	&  28/68.8  	&  56/108.9 	&  36/41.3  	& $3.6\times10^{30}$ 	& Opt.	        & CV/AB \\
    CX 5\ldots 	&  16 23 34.300 (4) 	& -26 29 56.47 (6) 	&  23/55.6  	&  35/66.7 	&  12/13.4  	& $2.5\times10^{30}$ 	& V56	        & amb.\\
    CX 6\ldots 	&  16 23 38.058 (9) 	& -26 29 21.81 (14) 	&  15/35.2  	&  26/47.8 	&  12/12.9  	& $1.8\times10^{30}$ 	& --	        & \\
    CX 7\ldots 	&  16 23 45.848 (12) 	& -26 28 55.03 (19) 	&  23/54.9  	&  25/46.3 	&   3/2.9   	& $1.8\times10^{30}$ 	& Opt.	        & fg.\\
    CX 8\ldots 	&  16 23 31.478 (9) 	& -26 30 57.84 (7) 	&  17/41.2  	&  24/46.0 	&   7/7.8   	& $1.7\times10^{30}$ 	& V52	        & AB \\
    CX 9\ldots 	&  16 23 34.224 (5) 	& -26 29 55.69 (5) 	&  17/41.0  	&  23/43.7 	&   6/6.7   	& $1.7\times10^{30}$ 	& V56	        & amb. \\
    CX 10\ldots	&  16 23 35.047 (8) 	& -26 31 19.18 (7) 	&  14/34.0  	&  23/43.9 	&   9/10.0  	& $1.7\times10^{30}$ 	& Opt.	        & AB\\
    CX 11\ldots	&  16 23 32.399 (7) 	& -26 30 45.63 (6) 	&   8/19.4  	&  18/34.7 	&  10/11.4  	& $1.3\times10^{30}$ 	& --	        & \\
    CX 12\ldots	&  16 23 38.205 (4) 	& -26 31 54.21 (6) 	&  16/39.3  	&  17/32.7 	&   2/2.2   	& $1.2\times10^{30}$ 	& Rad./Opt.	& MSP\\
    CX 13\ldots	&  16 23 34.326 (7) 	& -26 32 02.33 (10) 	&  11/26.9  	&  12/22.9 	&   2/2.1   	& $8.7\times10^{29}$ 	& Opt./V49	& AB\\
    CX 14\ldots	&  16 23 25.948 (15) 	& -26 33 54.62 (10) 	&   0/--    	&  12/23.1 	&  12/13.9  	& $8.7\times10^{29}$ 	& --	        & \\
    CX 15\ldots	&  16 23 36.769 (8) 	& -26 31 44.60 (11) 	&  10/24.6  	&  11/21.1 	&   1/1.0   	& $8.0\times10^{29}$ 	& Opt./V48	& AB\\
    CX 16\ldots	&  16 23 33.682 (6) 	& -26 34 17.27 (10) 	&   2/4.7   	&   6/11.4 	&   7/7.9   	& $4.3\times10^{29}$ 	& --    	& \\
    CX 17\ldots	&  16 23 35.975 (11) 	& -26 31 01.89 (8) 	&   7/16.9  	&   9/17.0 	&   2/2.2   	& $6.4\times10^{29}$ 	& --            & \\
    CX 18\ldots	&  16 23 45.767 (9) 	& -26 31 16.85 (14) 	&   7/17.0  	&   7/13.2 	&   0/--    	& $5.0\times10^{29}$ 	& Opt./V55 	& AB\\
    CX 19\ldots	&  16 23 28.953 (8) 	& -26 29 51.47 (9) 	&   4/9.5   	&   7/13.3 	&   3/3.3   	& $5.0\times10^{29}$ 	& --    	& \\
    CX 20\ldots	&  16 23 36.881 (11) 	& -26 31 39.43 (15) 	&   5/12.2  	&   7/13.3 	&   2/2.2   	& $5.0\times10^{29}$ 	& Opt.  	& AB\\
    CX 21\ldots	&  16 23 34.610 (5) 	& -26 32 04.60 (19) 	&   4/9.7   	&   7/13.3 	&   3/3.3   	& $5.0\times10^{29}$ 	& Opt.  	& \\
    CX 22\ldots	&  16 23 33.362 (9) 	& -26 31 45.65 (15) 	&   5/12.2  	&   6/11.4 	&   1/1.1   	& $4.3\times10^{29}$ 	& Opt.  	& AB\\
    CX 23\ldots	&  16 23 40.189 (20) 	& -26 29 26.10 (25) 	&   5/11.6  	&   6/10.8 	&   1/0.9   	& $4.1\times10^{29}$ 	& --    	& \\
    CX 24\ldots	&  16 23 42.096 (12) 	& -26 31 37.03 (13) 	&   3/7.2   	&   5/9.4	&  2/2.2    	& $3.6\times10^{29}$ 	& Opt.  	& amb.\\
    CX 25\ldots	&  16 23 33.503 (9) 	& -26 32 30.28 (8) 	&   4/10.0  	&   5/9.8	&  1/1.1    	& $3.7\times10^{29}$ 	& Opt.  	& AB\\
    CX 26\ldots	&  16 23 38.884 (11) 	& -26 31 48.26 (8) 	&   3/7.3   	&   5/9.5	&  2/2.2    	& $3.6\times10^{29}$ 	& Opt.  	& AB\\
    CX 27\ldots	&  16 23 33.290 (5) 	& -26 31 57.81 (25) 	&   4/9.7   	&   5/9.4  	&  1/1.0    	& $3.6\times10^{29}$ 	& Opt.  	& AB\\
    CX 28\ldots	&  16 23 34.969 (12) 	& -26 32 24.65 (10) 	&   4/9.9   	&   4/7.6  	&  0/--     	& $2.9\times10^{29}$ 	& Opt.  	& AB\\
    CX 29\ldots	&  16 23 19.486 (19) 	& -26 31 43.39 (26) 	&   1/2.3   	&   1/1.7  	&  2/2.3    	& $6.5\times10^{28}$ 	& --    	& \\
    CX 30\ldots	&  16 23 28.396 (9) 	& -26 30 22.25 (23) 	&   2/5.9   	&   3/7.0   	&   1/1.4   	& $2.6\times10^{29}$ 	& --    	& \\
    CX 31\ldots &  16 23 29.424 (9) 	& -26 30 00.19 (11) 	&   1/2.3   	&   1/1.7  	&  0/--     	& $6.5\times10^{28}$ 	& --    	& 
    \enddata
    \tablenotetext{a}{Sources are numbered according to their total counts.}
    \tablenotetext{b}{The \chandra\ positions have been corrected for
      a shift of $0\farcs15$ in declination (see
      \S\ref{ssec:opt_astro}). Positional uncertainties are given in
      parentheses and refer to the last quoted digit and are the
      centroiding uncertainties given by {\tool wavdetect}.  }
    \tablenotetext{c}{Corrections are described in
      Sect.~\ref{sec:cts}. The X-ray bands are 0.5--1.5~keV (\xsoft),
      0.5--4.5~keV (\xmed) and 1.5--6.0~keV (\xhard).}
    \tablenotetext{d}{For sources CX1--CX4, \Lx\ comes from an average
      of the best-fit models for each source
      (Sect.~\ref{sec:specfit}). A linear relation between \Lx\ and
      \xmed\ counts for these sources was derived and used to estimate
      \Lx\ for sources CX5--CX31 based on their \xmed\ counts. Typical
      uncertainties in \Lx\ are $\sim15$~\%.}
   \tablenotetext{e}{Type of counterpart (optical and/or radio) found
      and associations (if any) with previously reported sources. The
      ``V'' numbers refer to optical variables by Kaluzny et
      al.\,(1997\nocite{ktk97}) and Mochejska et
      al.\,(2002\nocite{mktp02}). R9 is an X-ray source detected by
      \rosat\ (Verbunt\,2001\nocite{ver01}).}
   \tablenotetext{f}{Classification of the sources; see
      Sect.~\ref{sec:classify}. The abbreviation MSP stands for
      millisecond pulsar, CV for cataclysmic variable and AB for
      active binary. Sources for which the classification is ambiguous
      are abbreviated with \emph{amb.} and the foreground source CX7
      is classified as \emph{fg.}.
      See text for details.}
  \end{deluxetable*}

  \subsection{Count Rates} \label{sec:cts}
  We extracted source counts in the following bands: 0.5--1.5~keV
  (\xsoft), 0.5--4.5~keV (\xmed), and 1.5--6.0~keV (\xhard).  The
  detected count rate was corrected for background, exposure variations,
  and foreground photoelectric absorption.  We make these corrections in
  order to produce an X-ray color-magnitude diagram (CMD) that can be
  compared to the X-ray CMDs that have resulted from
  \chandra\ observations of other globular clusters.  In addition,
  however, attention must be paid to differences in detector responses
  and, of course, exposure times and distances. 

  The background count rate in each band was estimated from a
  source-free region on the S3 chip outside the core and to the
  northwest.  The density of background counts in each band (for the
  25.8~ksec observation) is 0.0043 counts pixel$^{-1}$ in \xsoft, 0.0077
  counts pixel$^{-1}$ in \xmed, and 0.0047 counts pixel$^{-1}$ in
  \xhard.  The background count rate in the core may be somewhat higher,
  but even factors of a few greater than this estimate have negligible
  effects on our analysis.

  In general, the exposure variations among sources were at the
  $\sim$7\% level or less, but CX30 had an exposure which was
  20\% less than the others.  To account for these variations in
  exposure, we applied multiplicative corrections based on the ratio
  of the average effective area of the detector at the location of a
  source in each of the three bands to that in the same band of CX6,
  which had the highest average exposure.  The individual effective
  area curves for the sources were made using the CIAO tool {\tool
  mkarf}.  The average effective area of the detector at the location
  of CX6 in each of the bands was 538~\cmsq\ (\xsoft), 451~\cmsq\
  (\xmed), and 376~\cmsq\ (\xhard).

\begin{figure}
    \resizebox{85mm}{!}{\plotone{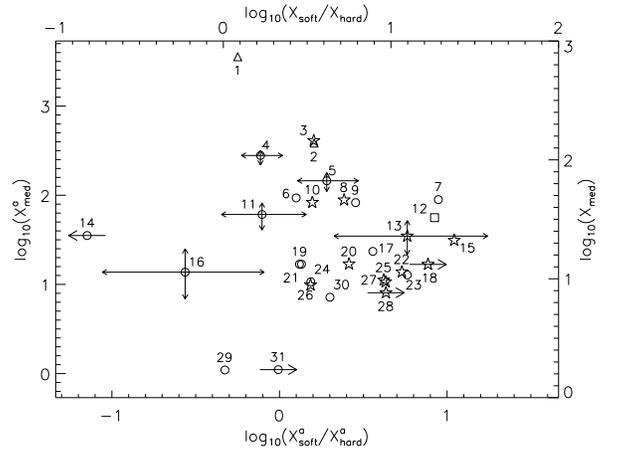}}
    \caption{X-ray color-magnitude diagram. The X-ray color is defined
      as the logarithm of the ratio of \xsoft\ (0.5--1.5~keV)
      corrected counts to \xhard\ (1.5--6.0~keV) corrected counts, and
      the magnitude is the logarithm of \xmed\ (0.5--4.5~keV)
      corrected counts.  Our correction for photoelectric absorption
      has the effect of uniformly shifting the data $+$0.27 units on
      the left axis and $+$0.34 units on the bottom axis. The bottom
      and left axes provide the absorbed color and magnitude scales
      ($^{\rm a}$), i.e., the observed colors and magnitudes
      uncorrected for absorption. For the sake of clarity, only a few
      error bars (1$\sigma$) are shown. Sources CX14, CX18, CX28 and
      CX31 were not detected in either the \xsoft\ or \xhard\ band,
      respectively. We illustrate their limits on $\log \xsoft /
      \xhard$ by adopting a single count in these bands. The X-ray
      sources are marked as in Fig.~\ref{fig:xrayimage}. See
      Section~\ref{sec:classify} for the source
      classification.
      \label{fig:xraycmd}}
  \end{figure}

  While the previous corrections were relatively minor (at the few
  percent level or less), the correction for photoelectric absorption
  is appreciable for M4.  The conversion of optical extinction
  to column density (Predehl \& Schmitt\,1995\nocite{ps95}) gives a
  value of $N_H=\ee{2.36}{21}~\pcmsq$.  We investigated the effects of
  such an absorption on three characteristic spectra: a 3~keV thermal
  bremsstrahlung, a 0.3~keV blackbody plus power law with photon index
  of $\Gamma=2$, and a power law with a photon index of $\Gamma=2$.
  The effects were most prominent in the \xsoft\ band, where the
  absorbed count rate was a factor of 2.2--2.5 lower than the
  unabsorbed one (depending on the spectrum).  Averaging the results
  of each spectrum in each band, we use the following correction
  factors: 2.38 (\xsoft), 1.87 (\xmed), and 1.10 (\xhard).
  Table~\ref{tab:srcs} lists both the observed and fully corrected
  counts in each band.  The effect of the absorption correction on the
  X-ray CMD (Fig.~\ref{fig:xraycmd}) is a uniform
  shift of all points 0.27 units on the left axis and 0.34 units on
  the bottom axis.  The bottom and left axes give the X-ray color and
  magnitude without this shift (they do, however, include the small
  corrections for background subtraction and exposure variations).

  \pagebreak[4]
  \subsection{Spectral Fitting}\label{sec:specfit}
  We used the CIAO tool {\tool dmextract} to extract spectra of
  sources CX1--CX4 in the 0.3--10~keV range.  We binned the spectra to
  have at least 10 counts (20 counts for CX1) per bin and fit them in
  \xspec\ (Arnaud\,1996\nocite{arn96}) using $\chi^2$ statistics.

  For CX2, CX3, and CX4, three different models (with absorption) were
  fit: thermal bremsstrahlung (TB), blackbody with a small
  ($\sim$20\%) power-law contribution (BB+PL), and power law (PL).  We
  fixed \nh\ to the value from optical extinction.  As expected for
  such low-count spectra, very few fits could be formally ruled out.
  We estimated the unabsorbed source luminosities by averaging the
  results from the three best-fit models for each source.  The spread
  in \Lx\ of the three models was $\sim$15\% for each source.  Fitting
  a linear relation to these luminosities versus corrected \xmed\
  counts, we have estimated the unabsorbed luminosities for sources
  CX5--CX31 based on their \xmed\ counts (note though that any
  differences between spectra are not accounted for). These are listed
  in Table~\ref{tab:srcs}.

\begin{figure}
    \resizebox{85mm}{!}{\plotone{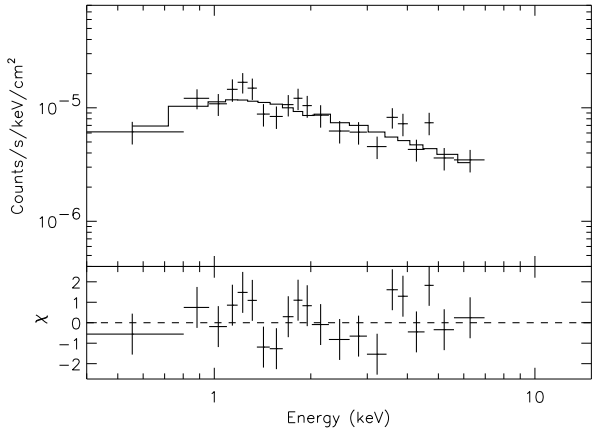}}
    \caption{\chandra\ spectrum of the brightest source, CX1. Datapoints are
      shown as crosses in the top panel, and an absorbed power-law model is
      shown as a solid line; residuals from this fit are shown in the bottom
      panel. Both the data and model have been divided by the instrument
      effective area for the purpose of plotting.\label{fig:cx1}}
  \end{figure}

  The X-ray color and magnitude of CX1 (Fig.~\ref{fig:xraycmd})
  suggested that it might be a CV, but the spectral fits using an
  absorbed TB model (with \nh\ allowed to vary) preferred the maximum
  temperature allowed by the model in \xspec, namely, $kT=200$~keV,
  which gave a $\chi^2$/d.o.f.\ of 36.0/39.  Fixing the temperature at
  the more reasonable value of $kT=25$~keV still allowed for a
  statistically acceptable fit ($\chi^2$/d.o.f.\ of 42.1/40).  The TB results
  indicated a rather hard spectrum, and an absorbed PL model (again with
  \nh\ allowed to vary) gave a best-fit photon index of $0.99\pm0.17$
  and best-fit \nh\ of \ee{(2.6\err{1.1}{0.7})}{21}~\pcmsq\ with a
  $\chi^2$/d.o.f. of 33.3/39.  The PL fit is shown in
  Fig.~\ref{fig:cx1}, and the unabsorbed luminosity from this model is
  \ee{6.8}{31}~\ergsec\ in the 0.5--2.5~keV band.  

  \section{Optical Observations}
  Three fields located inside the half-mass radius have been observed
  with the Wide Field Planetary Camera 2 (WFPC2) aboard the \hst\
  (\emph{HST}), and are shown in Fig.~\ref{fig:xrayimage}. Deep
  observations (dataset GO-5461) of three fields were obtained to
  determine the mass and luminosity functions of the main and white
  dwarf sequence (Richer et al.1995, 1997\nocite{rfi+95,rfi+97}) and
  two of these fields (see Fig.~\ref{fig:xrayimage}, gray outlines)
  fall within the cluster half-mass radius. These two fields were
  imaged for 11\,800~s in F336W (hereafter \filt{U}{336}), 15\,000~s
  in F555W (\filt{V}{555}) and 5\,500~s in F814W (\filt{I}{814}) and
  were re-observed in \filt{I}{814}, with exposure times of 5\,680~s,
  (GO-8153) approximately 5 years later. This allowed separation of
  field stars from the low luminosity cluster members through proper
  motion properties (Bedin et al.\,2001\nocite{bakp01}). Finally there
  exist shallow (249~s) \filt{V}{555} and (249~s) \filt{I}{814}
  observations (GO-6116) originally obtained to search for the optical
  counterpart of PSR~B1620$-$26. These observations
  (Fig.~\ref{fig:xrayimage}, black outline) cover nearly the entire
  region enclosed by the core radius. Of the 31 sources listed in
  Table~\ref{tab:srcs}, 18 coincide with the three \emph{HST}/WFPC2 fields.

  This section outlines the data reduction, photometry and astrometry
  of the \emph{HST}/WFPC2 images.

  \subsection{Data Reduction and photometry}\label{ssec:opt_photo}
  The single \emph{HST} images and the association
  products\footnote{\url{http://archive.eso.org/archive/hst/wfpc2\_asn/}}
  were obtained from the ESO archive. The single images were already
  calibrated, including full bias subtraction and flat-fielding. The
  association products consist of co-added, cosmic-ray cleaned images
  and association files, which contain the shifts between the single
  images. These shifts were used as input for the further reduction of
  the data, using HSTphot 1.1 (Dolphin\,2000a\nocite{dol00a}). This
  package is especially written for the reduction and photometry of
  \emph{HST} images.

  First, the HSTphot task \emph{mask} was used on each image to mask
  bad pixels and image defects. This ensures that these pixels are not
  used in the further analysis. The task \emph{crmask} was used to
  remove the cosmic ray hits, which are widely present in the separate
  images. For each filter the images were compared against each other
  to remove the cosmic ray events. The HSTphot task \emph{hotpixels}
  was then used to mask the known hot pixels in the WFPC2 detectors.

  To obtain photometry for each individual image we used the
  photometry task \emph{hstphot} of HSTphot 1.1. This task takes as
  input all available images in each filter and produces a master list
  of positions and magnitudes for each star found. The positions and
  magnitudes are determined by fitting a model point spread function
  (PSF), to each star. After all stars are fitted it calculates
  aperture corrections and corrects for the charge transfer efficiency
  effect. These corrections, together with the zero-points are
  described in (Dolphin\,2000b\nocite{dol00b}).

  \begin{figure*}
    \resizebox*{\textwidth}{!}{\plotone{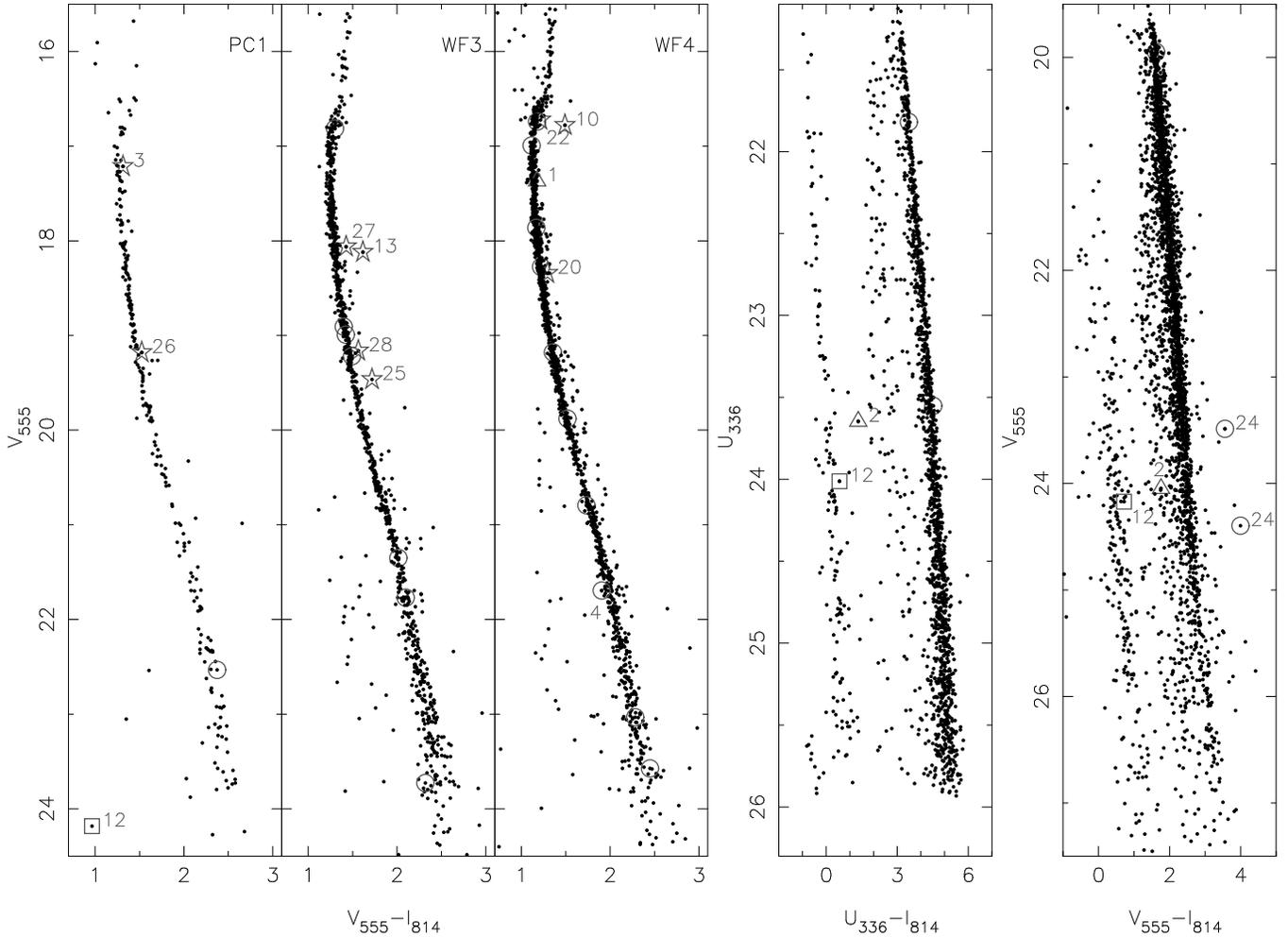}}
    \caption{Color-magnitude diagrams (CMDs) of the \emph{HST}/WFPC2
      observations of M4. The first three panels show CMDs from the
      GO-6116 dataset for the PC1, WF3 and WF4 chips, respectively (no
      X-ray sources coincide with the WF2 chip). The fourth and fifth
      panel show two CMDs of the data in both GO-5461 fields. All stars
      inside the 95\% confidence error circles are marked, while the
      candidate counterparts are numbered. The X-ray sources are marked
      as in Fig.~\ref{fig:xrayimage}. The candidate counterpart to CX24
      is plotted twice, as it was observed in both fields of the GO-5461
      dataset and had varied in brightness. The candidate counterpart
      was brightest during the observation of the field closest to the
      core of M4.\label{fig:cmds}}
  \end{figure*}

  \subsection{Astrometry}\label{ssec:opt_astro}
  To search for optical counterparts to the \chandra\ X-ray
  sources we aim to place both the X-ray and the optical frame onto
  the International Celestial Reference System (ICRS). We use this
  approach to improve the absolute pointing accuracy of \chandra\ and
  \emph{HST}, $0\farcs6$ and $1\farcs0$ ($1\sigma$) respectively
  (Aldcroft et al.\,2000\nocite{akc+00}; Biretta et
  al.\,2000\nocite{bir96}).

  To place the \emph{HST} images onto the ICRS we use two intermediate
  steps; first, we align a ground-based image onto the ICRS using an
  astrometric catalog, then we align the \emph{HST} images onto
  the ground-based image. To bring the X-ray frame onto the ICRS we use three
  optical identifications of X-ray sources.

  \subsubsection{Astrometry of the \emph{HST} images}
  A 2~minute $V$-band image, taken on February 21, 2002 with the Wide
  Field Imager (WFI) at the ESO 2.2~meter telescope on La Silla, was
  retrieved from the ESO archive and used to calibrate the
  \emph{HST}/WFPC2 images. The WFI has an array of 8 CCDs, each CCD
  having a $8\arcmin\times16\arcmin$ field of view, giving a total of
  $33\arcmin\times34\arcmin$. M4 is roughly centered on one
  chip, and, to minimize the effects of geometric distortion, an
  $8\arcmin\times8\arcmin$ sub-image, containing the entire area within
  the half-mass radius, was extracted from this chip.

  We found 115 stars on this sub-image that matched entries in the USNO
  CCD Astrograph Catalog (UCAC1, Zacharias et
  al.\,2000\nocite{zuz+00}). Of these stars, 91 were not saturated and
  appeared stellar and unblended and were used to compute an
  astrometric solution, fitting for zero-point position, scale and
  position angle. Five outliers, having residuals larger than
  $0\farcs2$ were rejected from the fit. The final solution has rms
  residuals of $0\farcs05$ in both coordinates.

  We calibrate the \emph{HST} frames with the astrometric solution of
  the WFI image. First, the pixel positions of the stars in the
  \emph{HST} datasets are placed on a single metaframe using the
  geometric distortion corrections and relative chip
  positions/orientations found by Anderson \&
  King\,(2003\nocite{ak03}). Next, using the astrometric solution in
  the FITS header of the \emph{HST} images, nominal celestial
  coordinates are computed for some 300--400 of the brightest
  stars. These positions are then matched with stars on the WFI image
  and their centroids are measured. Similar selection criteria for
  saturation and blending are used as with the UCAC stars. An
  astrometric solution is calculated from the calibrated celestial
  coordinates and the metaframe positions of the \emph{HST}
  stars. Outliers having residuals larger than three times the rms
  residual of the fit are removed, and a new solution is
  computed. This process is iterated until convergence. On average
  after convergence the astrometric solution contained some 200
  stars with rms residuals of $0\farcs07$ in both right
  ascension and declination.

  The 1$\sigma$ uncertainty in the optical positions is
  $0\farcs12$, the quadratic sum of the positional uncertainty
  in the tie of the WFI image to the UCAC ($0\farcs07$) and the
  uncertainty in the transfer to the \emph{HST} frames
  ($0\farcs09$) and the uncertainty in the tie of the UCAC onto
  the ICRS ($0\farcs02$, Assafin et al.\,2003\nocite{azrz+03}).

  By placing the \emph{HST} and WFI observations onto the ICRS with
  the UCAC we effectively set these images at the epoch of the UCAC
  observation, which is 1999.4. This implies that any average proper
  motion between the epochs of the \emph{HST} and WFI observations has
  been removed.

  The difference in proper motion between cluster members and
  background stars (Bedin et al.\,2003\nocite{bpka03}) possibly
  explains why our uncertainty in the calibration between \emph{HST}
  and WFI is larger than what we have found for similar calibrations
  of data from other clusters (e.g.\ Bassa et
  al.\,2003\nocite{bvkh03}).

  \subsubsection{Astrometry of the \chandra\ frame}
  We use three optical identifications in the \emph{HST} data of the
  \chandra\ X-ray sources to place the X-ray frame onto the ICRS frame
  of the \emph{HST}/WFPC2 images: the white dwarf companion of
  PSR~B1620$-$26 and two W~UMa variables.

  At the UCAC epoch, 1999.4, the position of PSR~B1620$-$26 is
  $\alpha_\mathrm{J2000}=16^\mathrm{h}23^\mathrm{m}38\fs2147(5)$,
  $\delta_\mathrm{J2000}=-26\degr31\arcmin53\farcs95(4)$, where we use
  the position and proper motion from Thorsett et
  al.\,(1999\nocite{tacl99}). According to our optical astrometry,
  this position is compatible with a white dwarf, at an offset of
  $-0\farcs04\pm0\farcs09$ ($-0\fs003\pm0\fs006$) in right
  ascension and $0\farcs07\pm0\farcs10$ in declination for the
  GO-5461 dataset. (The uncertainty in this offset is the quadratic
  sum of the uncertainty in the optical astrometry and the uncertainty
  in the corrected pulsar position.) With our accurate position
  ($0\farcs12$) for the white dwarf, we confirm the identification by
  Sigurdsson et al.\,(2003\nocite{srh+03}) and Richer et
  al.\,(2003\nocite{rifh03}) of this white dwarf as the companion of
  the pulsar. The position of X-ray source CX12 coincides with that of
  PSR~B1620$-$26.

  Five optical variables found by Kaluzny et al.\,(1997\nocite{ktk97})
  and Mochejska et al.\,(2002\nocite{mktp02}), roughly coincide with
  \chandra\ X-ray sources, three of which are in the \emph{HST}
  field(s) of view. Comparison of the Kaluzny et
  al.\,(1997\nocite{ktk97}) finding charts with our WFI and \emph{HST}
  images shows that the variables are indeed excellent matches to the
  X-ray sources.  The three variables coincident with the \emph{HST}
  images are all identified as W~UMa binaries. V55/CX18 is a single
  star on the \emph{HST} images, and we are confident that it is the
  star responsible for the optical variability. However, the variables
  V48 and V49 both are blends of four and two stars respectively. In
  the case of CX13/V49 one of the stars is clearly above the main
  sequence, indicating that this star is the variable. However, for
  CX15/V48 all stars of the blend lie on the main-sequence, and none
  of these can be securely identified as the variable.

  We therefore use the optical positions of the white dwarf companion
  to PSR~B1620$-$26 and the variables V49 and V55 to compute the shift
  needed to place the X-ray positions of CX12, CX13 and CX18 onto the
  optical positions. We have a total of 6 measurements, as CX12 is in
  the field of view of datasets GO-6116, GO-5461 and GO-8153, CX13
  only in that of GO-6116 and CX18 in those of GO-5461 and
  GO-8153. The weighted average shift is $0\farcs00\pm0\farcs05$ in
  right ascension and $0\farcs15\pm0\farcs05$ in declination, well
  within the $1\sigma$ uncertainty of the absolute \chandra\ pointing
  accuracy. We apply this shift as a boresight correction to the X-ray
  positions.

  \begin{figure*}
    \resizebox*{\textwidth}{!}{\plotone{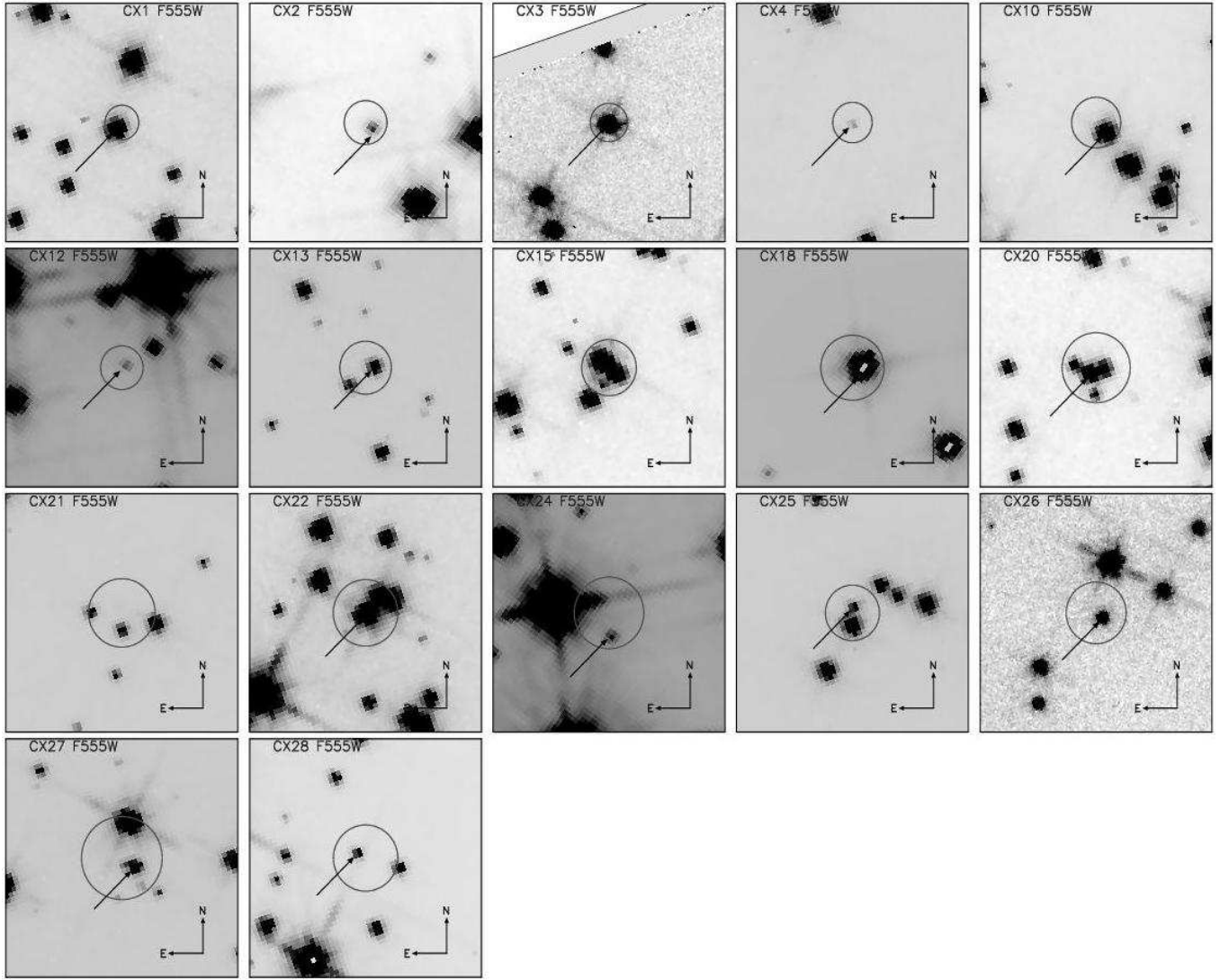}}
    \caption{$5\arcsec\times5\arcsec$ finding charts for the candidate
    optical counterparts. The finding charts are constructed from the
    coadded \filt{V}{555} ``association'' images. For sources CX2,
    CX12, CX18 and CX24, images from the GO-5461 dataset were used, images
    from the GO-6116 dataset are used for the other candidate
    counterparts. The 95\% confidence uncertainties on the
    \chandra\ positions are overlaid on these charts, while the
    candidate counterparts are indicated with an arrow. No arrow is
    present for sources CX15 and CX21 as no candidate counterparts
    were found. The pixel scale varies between sources coincident with
    WF chips and the PC1 chip (CX3 and CX26). The greyscale of these
    images is chosen as such to enhance the visibility of the
    candidate counterparts.\label{fig:charts}}
  \end{figure*}
  
  \pagebreak[4]
  \subsection{Identification of counterparts in the \emph{HST} images}
  We search for optical/UV stars inside 95\% confidence radii of the
  \chandra\ sources positions. The 1$\sigma$ uncertainty in a source
  position is the quadratic sum of the positional uncertainty for the
  X-ray source (Table~\ref{tab:srcs}), the uncertainty in the optical
  astrometry ($0\farcs12$) and the uncertainty in the X-ray
  boresight correction ($0\farcs07$). The 95\% confidence radius is a
  factor $[-2\log (1-0.95)]^{1/2}=2.448$ larger than the $1\sigma$
  uncertainty in the position.

  Each of the X-ray source error circles was visually checked for
  stars that were not found during the initial run of \emph{hstphot}. Stars
  that were missed were added by hand and the photometry task was
  executed again.

  The resulting photometry lists were used to create a set of
  color-magnitude diagrams (CMDs), shown in Figure~\ref{fig:cmds}.
  The data from the GO-6116 dataset is separated into three panels,
  one for each \emph{HST}/WFPC2 chip containing X-ray sources, as the
  cluster main sequence is displaced by about 0.1 magnitude in
  $\filt{V}{555}-\filt{I}{814}$ between the individual
  chips. Differential reddening towards M4 (Cudworth \&
  Rees\,1990\nocite{cr90}) is the likely cause for this. The data from
  the two GO-5461 fields, shown in the last two panels, did not
  display these displacements in color and the data from all four
  chips is plotted.

  We used the \emph{hstphot} information indicating the goodness of
  fit for a star to exclude the diffraction spike artifacts from the
  star list. These artifacts are widely present, especially in the
  GO-5461 dataset, which had the longest exposures. For more
  information we refer to the HSTphot 1.1
  manual\footnote{\url{http://www.noao.edu/staff/dolphin/hstphot/}}.
 
  All stars within the 95\% confidence radii of the \chandra\ source
  positions are marked in Figure~\ref{fig:cmds}. Stars with colors
  abnormal with respect to the cluster main sequence or giant branch
  are identified as candidate counterparts and are numbered in
  Fig.~\ref{fig:cmds}.  Positional and color information for each
  candidate counterpart is tabulated in Table~\ref{tab:opt}. Finding
  charts are shown in Figure~\ref{fig:charts}.

  \begin{deluxetable}{rrrrrlll}
    \tabletypesize{\footnotesize}
    \tablecolumns{8}
    \tablewidth{0pc}
    \tablecaption{Optical/UV Counterparts to \chandra\ X-ray Sources\label{tab:opt}}
    \tablehead{ 
      \colhead{CX} & \colhead{GO$^\mathrm{a}$} & \colhead{$\Delta \alpha$} & \colhead{$\Delta \delta$} &
      \colhead{$\Delta$} & \colhead{\filt{U}{336}} & \colhead{\filt{V}{555}} & \colhead{\filt{I}{814}} \\ 
      \colhead{} & \colhead{} & \colhead{($\arcsec$)} & \colhead{($\arcsec$)} &
      \colhead{($\sigma$)} & \colhead{} & \colhead{} & \colhead{} }
    \startdata
    1  & 6116 &  0.24 & -0.14 & 2.0 & \ldots     & $17.37(1)$ & $16.19(1)$ \\[0.5em]
    2  & 5461 & -0.02 & -0.19 & 1.2 & $23.65(6)$ & $24.05(3)$ & $22.29(3)$ \\
       & 8153 &  0.08 & -0.07 & 0.7 & \ldots     & \ldots     & $22.04(1)$ \\[0.5em]
    3  & 6116 &  0.12 & -0.00 & 0.8 & \ldots     & $17.22(1)$ & $15.90(1)$ \\
       & 5461 &  0.19 & -0.05 & 1.3 & $18.00(1)$ & sat.       & sat.       \\
       & 5461 &  0.16 &  0.04 & 1.1 & $17.94(1)$ & sat.       & sat.       \\
       & 8153 &  0.20 &  0.02 & 1.4 & \ldots     & \ldots     & $15.94(1)$ \\
       & 8153 &  0.05 &  0.06 & 0.6 & \ldots     & \ldots     & $18.94(1)^\mathrm{b}$ \\[0.5em]
    4  & 6116 &  0.12 & -0.01 & 0.8 & \ldots     & $21.69(3)$ & $19.79(2)$ \\[0.5em]
    10 & 6116 & -0.09 & -0.21 & 1.2 & \ldots     & $16.78(1)$ & $15.29(1)$ \\[0.5em]
    12 & 6116 &  0.02 & -0.00 & 0.1 & \ldots     & $24.2(2)$  & $23.2(2)$  \\
       & 5461 &  0.08 &  0.03 & 0.5 & $24.01(8)$ & $24.17(2)$ & $23.45(4)$ \\
       & 8153 &  0.11 &  0.02 & 0.7 & \ldots     & \ldots     & $23.49(3)$ \\[0.5em]
    13 & 6116 & -0.08 &  0.03 & 0.4 & \ldots     & $18.11(1)$ & $16.51(1)$ \\[0.5em]
    18 & 5461 & -0.09 & -0.05 & 0.4 & $17.76(1)$ & sat.       & sat.       \\
       & 8153 & -0.03 & -0.04 & 0.2 & \ldots     & \ldots     & $15.64(1)$ \\[0.5em]
    20 & 6116 &  0.24 & -0.13 & 1.1 & \ldots     & $18.34(1)$ & $17.05(1)$ \\
       & 5461 &  0.25 & -0.15 & 1.1 & $19.62(1)$ & sat.       & sat.       \\
       & 8153 &  0.26 & -0.08 & 1.1 & \ldots     & \ldots     & $17.04(1)$ \\[0.5em]
    22 & 6116 &  0.11 & -0.03 & 0.5 & \ldots     & $16.73(1)$ & $15.50(2)$ \\[0.5em]
    24 & 5461 &  0.08 & -0.46 & 1.8 & und.       & $23.49(1)$ & $19.93(1)$ \\
       & 5461 &  0.14 & -0.52 & 2.1 & und.       & $24.4(1)$  & $20.40(3)$ \\
       & 8153 &  0.23 & -0.38 & 1.8 & \ldots     & \ldots     & $19.93(1)$ \\
       & 8153 &  0.20 & -0.38 & 1.7 & \ldots     & \ldots     & $20.02(1)$ \\[0.5em]
    25 & 6116 &  0.09 &  0.14 & 0.8 & \ldots     & $19.45(3)$ & $17.75(1)$ \\[0.5em]
    26 & 6116 & -0.03 & -0.10 & 0.5 & \ldots     & $19.18(1)$ & $17.66(1)$ \\
       & 5461 & \ldots & \ldots & \ldots & sat.       & sat.       & sat.       \\
       & 5461 &  0.02 & -0.05 & 0.2 & $20.23(1)$ & sat.       & sat.       \\
       & 8153 &  0.06 & -0.09 & 0.5 & \ldots     & \ldots     & $17.83(1)$ \\
       & 8153 &  0.06 & -0.09 & 0.5 & \ldots     & \ldots     & $17.76(2)$ \\[0.5em]
    27 & 6116 & -0.16 & -0.18 & 0.8 & \ldots     & $18.06(1)$ & $16.65(1)$ \\[0.5em]
    28 & 6116 &  0.29 &  0.11 & 1.3 & \ldots     & $19.16(1)$ & $17.61(1)$ \enddata
    \tablecomments{Optical positions and magnitudes of the candidate
      counterparts. The optical positional is given by the
      \chandra\ value plus offset (in true seconds of arc) in
      each coordinate. Magnitude uncertainties are given in
      parentheses and refer to the last quoted digit. Stars that were
      saturated in a pass band are denoted with ``sat.'' and stars
      that are not detected in a band are denoted with ``und.''.}
    \tablenotetext{a}{The \emph{HST} GO dataset number from which the
      position and magnitudes are determined. For CX3, CX24 and CX26
      the first entry for dataset GO5461 and GO8153 refers to the
      field nearest to the cluster center.}
    \tablenotetext{b}{This measurement is likely in error due to the
      proximity of the star to the edge of the chip.}
  \end{deluxetable}

  \subsection{Identification of counterparts in the WFI image}
  Several of \chandra\ X-ray sources coincide with stars on the WFI
  image. Two of these stars, identified by Mochejska et
  al.\,(2002\nocite{mktp02}) as variables V52 and V56 are coincident
  with X-ray sources CX8 and the possibly blended source CX5-9,
  respectively. The offset of the optical positions with respect to
  the \chandra\ positions in Table~\ref{tab:srcs} is
  $0\farcs06\pm0\farcs20$, $0\farcs22\pm0\farcs21$ for CX8 and V52,
  $-0\farcs29\pm0\farcs22$, $-0\farcs23\pm0\farcs21$ for CX5 and V56
  and $0\farcs73\pm0\farcs20$, $0\farcs55\pm0\farcs21$ for CX9 and
  V56.

  The X-ray source CX7 coincides with the UCAC1 star 22560073, a
  foreground star according to its proper motion (Cudworth \&
  Rees\,1990\nocite{cr90}, star A330 with $V=12.68$).  The offset of
  the corrected X-ray position, as listed in Table~\ref{tab:srcs},
  from the optical position from UCAC is $0\farcs38\pm0\farcs18$ in
  right ascension and $-0\farcs16\pm0\farcs20$ in declination. The
  error in this offset is dominated by the uncertainty in the X-ray
  position of CX7 in the \chandra\ frame; much smaller contributions
  to the error are due to the uncertainties in the X-ray to optical
  shift and in the optical position. Redetermining the X-ray to
  optical shift including CX7 with the other three optically
  identified X-ray sources, does not significantly change the value of
  the shift.

  \section{Source Classification}\label{sec:classify}
  In attempting to classify the X-ray sources detected in M4, we note
  that five of our optical counterparts to \chandra\ sources have a
  previously known classification.  CX12 is a millisecond pulsar with
  a white dwarf companion (Thorsett et al.\,1999); the optical
  counterpart corresponds to the white dwarf (Sigurdsson et al.\ 2003,
  see also Fig.~\ref{fig:cmds}).  CX13/V49, CX15/V48 and CX18/V55 are
  contact binaries, and CX8/V52 is a BY~Dra system, i.e.\ these four
  sources are all magnetically active binaries (Kaluzny et
  al.\,1997\nocite{ktk97}; Mochejska et al.\,2002\nocite{mktp02}). The
  optical counterparts to CX13 and CX15 as identified from the ground
  based data, have both been resolved into multiple objects with
  \emph{HST}/WFPC2. Kaluzny et al.\,(1997\nocite{ktk97}) noted that
  the optical counterpart to CX13, was brighter than expected for a
  contact binary with its period, placed at the distance and reddening
  of M4. This discrepancy is removed as our counterpart to CX13 is
  about a magnitude fainter than the blend (we note that the optical
  counterpart to CX13 appears to be too bright to be on the binary
  main sequence as it is about a magnitude brighter than a star on the
  main sequence with the same color, Figure~\ref{fig:cmds}). For the
  optical counterpart of CX15 the opposite appears to be the case. It
  was seen as a cluster member by Kaluzny et
  al.\,(1997\nocite{ktk97}), but \emph{HST}/WFPC2 observations
  indicate that V48 is a blend of 3 or 4 stars. Hence, the contact
  binary is too faint to be a member, and may be a background object.

  The possible blend of CX5 and CX9 coincides with the variable giant
  V56 found by Mochejska et al.\,(2002\nocite{mktp02}).  If the giant
  is a member of a RS~CVn binary then its variability and X-ray
  emission can be explained by magnetic activity. However, without
  more information on the optical properties of V56 the classification
  remains ambiguous.

  \begin{figure}
    \resizebox{85mm}{!}{\plotone{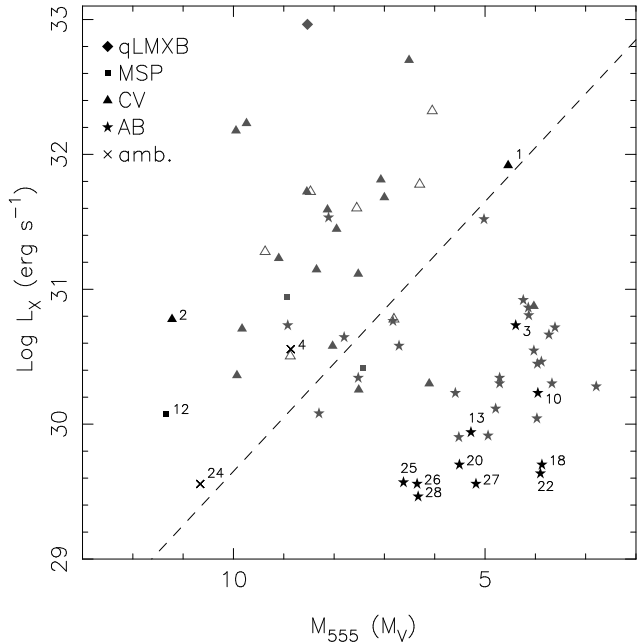}}
    \caption{The X-ray luminosity \Lx\ (0.5--2.5~keV) and the absolute
    \filt{V}{555} (approximately $V$-band) magnitude for X-ray sources
    with identified optical counterparts. Five types of X-ray sources
    are shown, quiescent LMXBs (\emph{diamonds}), MSPs
    (\emph{squares}), CVs (\emph{triangles}), ABs (\emph{stars}) and
    unclassified (\emph{crosses}). The closed points are sources from
    47\,Tuc (Grindlay et al.\,2001, Edmonds et al.\,2003a) and the
    open points are from NGC\,6752 (Pooley et al.\,2002). The numbered
    points are the optical counterparts to the \chandra\ X-ray sources
    in M4, the number corresponds to the source number. The absolute
    magnitude of the optical counterparts is computed from the
    \filt{V}{555} magnitudes and the $V$-band distance modulus as in
    Harris\,(1996) for 47\,Tuc and NGC\,6752 and that of Richer et
    al.\,(1997) for M4. The dashed line of constant X-ray to optical
    flux ratio roughly separates cataclysmic variables from active
    binaries.\label{fig:lxlopt}}
  \end{figure}

  In classifying the remaining X-ray sources we first look at the
  X-ray emission itself. A neutron star accreting at a low rate from a
  companion star, i.e.\ a low-luminosity (quiescent) low-mass X-ray
  binary (qLMXB), is characterized by a soft X-ray spectrum (black
  body color temperature $\lesssim 0.3$~keV) and a luminosity
  $\Lx\gtrsim10^{32}$~\ergsec\ (Verbunt et al.\ 1994\nocite{vbj+94};
  Rutledge et al.\ 1999\nocite{rbb+99}).  None of the X-ray sources in
  our sample shows this characteristic, and we conclude that M4 does
  not contain a low-luminosity low-mass X-ray binary with an accreting
  neutron star.

  We extract further information from the location of the optical star
  in the color-magnitude diagrams of Fig.~\ref{fig:cmds}: stars bluer
  than the main-sequence are possible cataclysmic variables, stars
  above the main-sequence possible magnetically active binaries. The
  ratio of X-ray to optical flux is also useful, as it roughly
  separates the cataclysmic variables from magnetically active
  binaries (Verbunt \&\ Johnston 2000\nocite{vj00}; Pooley et al.\
  2002a\nocite{plh+02}). We illustrate this in Fig.~\ref{fig:lxlopt},
  where we plot data from 47\,Tuc (Edmonds et al.\
  2003a,b\nocite{eghg03a,eghg03b}) and NGC\,6752 (Pooley et al.\
  2002a\nocite{plh+02}), and add our data from M4.

  CX2 is a probable cataclysmic variable, because it is blue
  (Fig.~\ref{fig:cmds}) and has a relatively high X-ray to optical
  flux ratio (Fig.~\ref{fig:lxlopt}). High X-ray to optical flux
  ratios suggest that CX1 and CX4 also are cataclysmic variables, even
  though they are located on or close to the main sequence in the
  color-magnitude diagram (Fig.~\ref{fig:cmds}).

  The X-ray luminosity of CX1 is too high for a magnetically active
  binary of two main sequence stars (i.e.\, a BY Dra system), and its
  optical magnitude excludes that it is a magnetically active binary
  with a sub-giant (i.e.\, a RS CVn system). This, and the relative
  hardness of the X-ray spectrum (Fig.~\ref{fig:cx1}) indicate that it
  is a cataclysmic variable. We note that the optical counterpart to
  CX1 is offset from the \chandra\ position by $2\sigma$ and might be
  a chance coincidence. This would indicate that the actual optical
  counterpart is even fainter, possibly lost in the glare of the
  bright star, leading to higher X-ray to optical flux ratios, as seen
  with most other CVs.
  
  The less luminous counterpart to CX4 is probably a cataclysmic
  variable; its location near the main-sequence in
  $\filt{V}{555}-\filt{I}{814}$ has precedents in other globular
  clusters (Edmonds et al.\,2003a,b\nocite{eghg03a,eghg03b}).

  The optical counterpart of CX24 varies by more than a magnitude in
  brightness (Fig.~\ref{fig:cmds}). It is located further above or to
  the right of the main-sequence than the binary sequence. Hence we
  believe that CX24 is either a fore or background object and
  unrelated to M4. The same argument could be made for CX10. However,
  stars at similar positions in the color-magnitude diagram of other
  old clusters are X-ray sources and confirmed members (albeit
  unexplained, see e.g.\ Mathieu et al.\,2003\nocite{mbt+03} for M67
  and Orosz \& van Kerkwijk\,2003\nocite{ok03} for NGC\,6397) or
  probable members (Albrow et al.\,2001\nocite{agb+01}, Edmonds et
  al.\,2003a\nocite{eghg03a} for 47\,Tuc). We therefore consider CX10
  a probable cluster member.

  Based on positional coincidence we identify CX7 with the UCAC1 star
  22560073. This star has $V=12.68$, $B-V=0.88$ (Cudworth \&
  Rees\,1990\nocite{cr90}), which is compatible with a K2V star at a
  distance of about 200~pc. Scaling the X-ray luminosity from
  Table~\ref{tab:srcs} from the distance of M4 to a distance of 200~pc
  gives $\Lx\approx2.4\times10^{28}~\ergsec$, well within the range
  observed for ordinary K2V stars (Verbunt\,2001\nocite{ver01}).

  The relatively low X-ray to optical flux ratios of the remaining
  optical counterparts found with \emph{HST}/WFPC2 suggests that they
  are active binaries. The counterparts of CX13 and CX18, already
  known to be contact binaries from ground based data, are among
  these. (The $\filt{V}{555}$ image of the optical counterpart of CX18
  is over-exposed; we use the estimate
  $\filt{V}{555}=(\filt{U}{336}+\filt{I}{814})/2$.)  The others (CX3,
  CX20, CX22, CX25, CX26, CX27, CX28) are at or above the
  main-sequence, as expected for binaries, and we conclude that all of
  them are magnetically active binaries. The X-ray-to-optical flux
  ratio of any of the 2 candidate counterparts of CX21 is comparable
  to that of an active binary.

  \section{Discussion}
  The X-ray luminosities of the \chandra\ sources in M4
  are amongst the lowest ever observed in a globular cluster. 
  More than half of the X-ray sources in this cluster have
  $\Lx<10^{30}~\ergsec$, compared to 10\% of the sources in
  47\,Tuc (Grindlay et al.\,2001a\nocite{ghem01}) and 28\% of the
  sources in NGC\,6397 (Grindlay et al.\,2001b\nocite{ghe+01}).

  It is therefore not surprising that only one of the X-ray sources
  detected with \chandra\ may have been detected before. The
  position of the marginal \rosat\ HRI source R9 (Verbunt
  2001\nocite{ver01}) is 7$\farcs$5 from that of CX1. Even though this
  is further than expected from the error given for the \rosat\ HRI source,
  we think the identification is probable. The X-ray luminosity of the
  \rosat\ HRI source is only \ee{1.3}{31}\ \ergsec, a factor six
  below the luminosity detected with \chandra. The upper limit from the 
  \rosat\ PSPC observation is  \ee{1.5}{31}\ \ergsec. (These luminosities
  are re-computed for a distance of 1.73\,kpc and for the spectrum of
  CX1, and thus differ from those given by Verbunt; mainly because
  he used a distance of 2.2\,kpc.)  This large variability indicates
  that CX1 is a cataclysmic variable rather than a radio pulsar.
  Note that even if the \rosat\ source is not identical to CX1, it
  still provides an upper limit to the flux of CX1 during the
  \rosat\ observations, and thus proves that CX1 is highly variable.
  
  Several other X-ray sources coincide with optical variables discovered by
  Kaluzny et al.\,(1997\nocite{ktk97}) and Mochejska et
  al.\,(2002\nocite{mktp02}). The sources CX13, CX15 and CX18 
  are coincident with W~UMa variables V49, V48 and V55, with orbital
  periods of  0.283, 0.298 and 0.311~days, respectively.
  The upper limit on the X-ray luminosity of the other
  W~UMa binaries found by Kaluzny et al.\,(1997\nocite{ktk97}),
  V44, V47, V50, V51, V53 and V54, is about \ee{3}{29}\,\ergsec.
  Both detections and upper limits of these W UMa variables are
  in agreement with the range of X-ray luminosities of W~UMa
  binaries in the \rosat\ All Sky Survey (St\c{e}pie\'n, Schmitt \&
  Voges\,2001\nocite{ssv01}).

  CX8 coincides with the variable V52, which is classified as a BY~Dra
  system with a period of 0.777~days. The possible blend of X-ray
  sources CX5 and CX9 coincides with the variable V56. No period is
  known for this object, but its variability (a rise of 0.1 mag in 4
  days) and its location on the giant branch (Mochejska et
  al.\,2002\nocite{mktp02}) are suggestive of an RS CVn variable.  If
  it is, it is the first RS CVn binary detected in X-rays in a
  globular cluster -- a marked contrast with old open clusters, where
  RS CVns dominate the X-ray sources (Belloni et al.\
  1998\nocite{bvm98}). The exposure of the eclipsing binary V54 is not
  as good as for the other sources, hence the upper limit on the X-ray
  luminosity of this source is somewhat higher. Detections and upper
  limits of these binaries are well within the range observed for
  magnetically active binaries, e.g.\ in the \rosat\ All Sky Survey
  (Dempsey et al.\ 1993\nocite{dlfs93}).
  
  \begin{deluxetable}{lcccccc}
    \footnotesize 
    \tablecaption{Scaling Parameters of M4, NGC\,6397 and
    47~Tuc\label{tab:scaling}}
    \tablehead{
      \colhead{Cluster} & 
      \colhead{log $\rho_0$} &
      \colhead{$r_\mathrm{c}$} &
      \colhead{$d$} &
      \colhead{$M_V$} &
      \colhead{$\Gamma$} &
      \colhead{$M_\mathrm{c}$} \\
      \colhead{} &
      \colhead{($L_\sun\ \mathrm{pc}^{-3}$)} &
      \colhead{(\arcsec)} &
      \colhead{(kpc)} &
      \colhead{} &
      \colhead{} &
      \colhead{}}
    \startdata
    M4        & 4.01 & 49.8 & 1.73 & $-$6.9 & 1.0 & 1.0 \\
    NGC\,6397 & 5.68 &  3.0 & 2.3 & $-$6.6 & 2.1 & 0.024 \\
    47~Tuc    & 4.81 & 24.0 & 4.5 & $-$9.4 & 24.9 & 12.4 \enddata
    \tablecomments{Values for central density ($\rho_0$), core-radius
      ($r_\mathrm{c}$), distance ($d$) and absolute visual magnitude
      ($M_V$) originate from Harris\,1996 (version of February
      2003). For M4, the values of $\rho_0$ and $M_V$ are computed for
      the distance and reddening of Richer et al.\,(1997). The
      collision number is computed from $\Gamma \propto \rho_0^{1.5}\
      r_\mathrm{c}^2$ and the core mass from $M_\mathrm{c} \propto
      \rho_0\ r_\mathrm{c}^3$. Values for $\Gamma$ and $M_\mathrm{c}$
      are normalized to the value of M4.}
  \end{deluxetable}

  In trying to determine the X-ray luminosity function of M4, we can
  use the number of detected sources as a function of
  luminosity. However, the number of excess counts, not allocated to
  individually detected sources, also contains information about the
  luminosity function. If we assume a luminosity function
  $dN\propto{\Lx}^{-\gamma}d\log \Lx$ and a reference luminosity
  $L_\mathrm{r}$, then the ratio of the contributions to the total
  luminosity by sources in the ranges (0.1--1)$L_\mathrm{r}$ and
  (1--10)$L_\mathrm{r}$ is given by
  \begin{equation}
    R_L(L_\mathrm{r}) \equiv
    {\int_{0.1L_\mathrm{r}}^{L_\mathrm{r}}\Lx \mathrm{d} N\over\int_{L_\mathrm{r}}^{10L_\mathrm{r}}\Lx
    \mathrm{d} N} = 10^{\gamma-1} 
    \label{eqL}
  \end{equation}
  and the ratio of the number of sources in the same ranges is
  \begin{equation}
    R_N(L_\mathrm{r}) = 10^{\gamma} = 10 R_L(L_\mathrm{r}).
    \label{eqN}
  \end{equation}
  In particular, we note that for $\gamma<0.7$ and a luminosity
  function continuing to arbitrarily low luminosities the total
  luminosity of all sources with $\Lx < L_\mathrm{r}$ is less than the
  luminosity for the sources with (1--10)$L_\mathrm{r}$.  In the
  following we assume that the ratio of counts is proportional to the
  ratio of luminosities, and we use as a reference luminosity
  $L_\mathrm{r}\equiv6\times10^{29}$~\ergsec.

  Following the method described in Johnston \&
  Verbunt\,(1996\nocite{jv96}) and Pooley et
  al.\,(2002b\nocite{plv+02}) we derive from the list of detected
  X-ray sources (Table~\ref{tab:srcs}) that $\gamma=0.71$ for X-ray
  luminosities above $\Lx(0.5-2.5)=4.4\times10^{29}$~\ergsec. The K-S
  probability of this value is 93\%. For K-S probabilities above 10\%
  the slope of the luminosity function has $0.47<\gamma<1.07$.

  In addition to the detected sources, we have an excess in the core
  of M4 of about 150 counts, uncorrected for absorption, which
  corresponds to roughly 300 counts corrected for absorption in the
  0.5--6.0~keV band (which is the addition of counts in \xsoft\ and in
  \xhard). In this band, the five faintest sources in the core, CX20,
  CX21, CX22, CX26 and CX27 have between 14 and 10 counts. A minimum
  number of sources required to explain the excess counts is found by
  assuming that each source has 10--14 counts, which gives 22--30
  sources.

  The total number of $\sim\!350$ counts from sources {\em in the
  core} of M4 with $\Lx<L_\mathrm{r}$ (five detected sources and the
  excess) is similar to that of the 6 sources in the range
  (1--10)$L_\mathrm{r}$. From our remark following Eqs.\,\ref{eqL} and
  \ref{eqN}, we see that this implies that $\gamma\ge0.7$; for lower
  values of $\gamma$ the sources with $\Lx<L_\mathrm{r}$ do not
  contribute enough counts to explain the observed number. On the
  other hand, for $\gamma=1$ the luminosity function must have a
  cutoff near 0.1$L_\mathrm{r}$, because otherwise the sources with
  $\Lx<L_\mathrm{r}$ would produce more counts than is observed. With
  6 sources in the range (1--10) $L_\mathrm{r}$, and for
  $\gamma=0.7-1.0$ we have 30--60 core sources in the range
  (0.1--1)$L_\mathrm{r}$.

  In the core of 47\,Tuc there are 14 sources in the range (10--100)
  $L_\mathrm{r}$, which for $\gamma=0.7-0.8$ leads to predicted
  numbers of 70--90 (of which 27 are already detected individually) in
  the range (1--10) $L_\mathrm{r}$, and 350--560 in the range (0.1--1)
  $L_\mathrm{r}$.  These numbers are compatible with the total number
  of counts in the core of 47\,Tuc from the observed sources plus the
  excess of about 500 counts, estimated by Grindlay et al.\
  (2001a\nocite{ghem01}). 

  In Table~\ref{tab:scaling} we compare the collision numbers and the
  masses of the cores of M4 and 47\,Tuc.  We expect that the numbers
  of cataclysmic variables scales with the collision number, and the
  number of magnetically active binaries with the mass. In the range
  (1--10) $L_\mathrm{r}$ the ratio of numbers of sources in the cores
  of M4 and 47\,Tuc suggest that the ratio is set by mass rather than
  collision number. This in turn suggests that magnetically active
  binaries dominate the numbers at these luminosities, in agreement
  with our suggested identifications in M4.

  Comparison with NGC\,6397 is also interesting, as its collision
  number is somewhat higher, $\Gamma=2.1$ (normalized to the value of
  M4) while its total mass (as measured from its absolute magnitude)
  is somewhat lower than that of M4. Within its half-mass radius,
  NGC\,6397 contains 9 cataclysmic variables, 4 magnetically active
  binaries and 10 as yet unclassified sources (generally of low X-ray
  luminosity). Within the half mass radius of M4 we find 2 or 3
  cataclysmic variables, 12 magnetically active binaries and 15
  unclassified sources, of which an estimated 6--12 are background
  sources. The ratio for the numbers in NGC\,6397 and M4 of
  magnetically active binaries is as expected for a scaling with
  mass. As already noted by Pooley et al.\ (2003\nocite{pla+03}),
  NGC\,6397 contains rather more cataclysmic variables than expected
  on the basis of its collision number.  This is reflected in the flat
  slope of its X-ray luminosity function. If the high number of bright
  systems in NGC\,6397 is explained by a higher mass of this cluster
  in the past (as suggested by Pooley et al.\ 2003\nocite{pla+03}),
  one would expect an accordingly higher number of magnetically active
  binaries, in contrast to what is observed (assuming that
  magnetically active binaries have a similar evaporation rate as
  cataclysmic variables, much lower than the evaporation rate of
  single stars).  In this respect, it is worth noting that the time
  scale on which a binary can be destroyed by a close encounter is
  short in NGC\,6397 compared to most other clusters, including M4 and
  47\,Tuc (Verbunt 2003\nocite{ver03}, in particular Figure 3b). Since
  magnetically active binaries tend to have longer orbital periods
  then cataclysmic variables, they would be easier destroyed by close
  encounters. This might explain the absence of large numbers of such
  binaries in NGC\,6397.

  \acknowledgements C.B. acknowledges support by the Netherlands
  orginization for Scientific Research (NWO). D.P. and W.H.G.L. thank
  NASA for its support. L.H. and S.F.A. also gratefully acknowledge
  their support from NASA grant NAG5-7932. V.M.K. is supported by
  NSERC, NATEQ, CIAR and NASA. This research is based on observations
  made with the ESO Telescopes at the La Silla Observatory and
  observations made with the NASA/ESA Hubble Space Telescope, obtained
  from the data archive at the Space Telescope Institute. STScI is
  operated by the association of Universities for Research in
  Astronomy, Inc. under the NASA contract NAS 5-26555. Support for
  proposal \#9959 was provided by NASA through grants from STScI.

\end{document}